\newcommand{\ignore}[1]{}

\documentstyle[sprocl]{article}

\input{psfig.sty}

\def\be{\begin{equation}}
\def\ee{\end{equation}}
\def\bea{\begin{eqnarray}}
\def\eea{\end{eqnarray}}

\newcommand{\text}[1]{\qquad \mbox{#1} \qquad}

\begin{document}

\title{ FROM BLACK HOLES TO POMERON: \\
Tensor Glueball and Pomeron Intercept at Strong Coupling\footnote{ Talk 
presented by R. C. Brower at ISMD99: QCD and Multiuparticle Production.
 This work was supported in part by the Department
of Energy under Contract No. DE-FG02/91ER40688 and
DE-FG02-91ER40676}}
\author{Richard  C. Brower\address{Physics Department,
        Boston University, 590 Commonwealth Ave, Boston, MA 02215, USA },
      Samir D. Mathur\address{Center for Theoretical
        Physics, Mass. Inst. of Technology, Cambridge, MA 02139, USA } 
        and Chung-I Tan\address{Physics Department, 
        Brown University, Providence, RI 02912, USA}}


\maketitle\abstracts{We briefly review the approach for strong coupling 
calculation of glueball masses
based on the duality between supergravity and Yang-Mills theory.  Earlier work 
is extended to non-zero spin. Fluctuations in
the gravitational metric lead to the $2^{++}$ tensor glueball state on the 
leading
Pomeron trajectory with a mass relation: $m(0^{++}) < m(2^{++}) $. 
In particular, for
$QCD_4$, a strong coupling expansion for the Pomeron intercept is
obtained. }

\section{Introduction}

The Maldacena conjecture~\cite{maldacena} and its further extensions
allow us to compute quantities in a strongly coupled gauge theory from
its dual gravity description. In particular, Witten~\cite{wittenT}~
has pointed out if we compactify the 4-dimensional conformal super
Yang Mills (SYM) to 3 dimensions using anti-periodic boundary
conditions on the fermions, then we break supersymmetry and conformal
invariance and obtain a theory that has interesting mass scales. 
This
approach has been used to calculate a discrete mass spectrum for
${\tilde 0}^{++}$ states associated with $Tr[F^2]$ at strong coupling
by solving the dilaton's wave equation in the corresponding gravity
description.\cite{csaki,jev} Although the theory at strong coupling is
really not pure Yang-Mills, since it has additional fields, some rough
agreement was claimed with the pattern of glueball masses.

Here we report on the calculation of the discrete modes for the
perturbations of the gravitational metric.\cite{bmt,cm} A complete
description for all discrete fluctuations has also been carried out, both
for $QCD_3$ and $QCD_4$.\cite{bmt4}  For simplicity, we shall discuss here 
mostly  $QCD_3$. 
For $QCD_4$,  from the mass  of the $2^{++}$ state
and the calculated QCD string tension, we obtain a strong coupling expansion
for the Pomeron intercept: $\alpha_P(0)= 2- 0({1/g^2N})$. In this approach, the 
Pomeron corresponds to a
``massive graviton".  Other 
results will be reported
elsewhere.\cite{bmt4}

\section{AdS/CFT Duality at Finite $\beta$}
Let us review briefly the proposal for getting a 3-d Yang-Mills theory
dual to supergravity. One begins by considering Type IIB supergravity
in Euclidean 10-dimensional spacetime with the topology 
$M_5\times S^5$.  The  Maldacena conjecture asserts that
IIB superstring theory on $AdS^5\times S^5$ is dual to the
${\cal N } = 4$ SYM conformal field theory on the boundary of the $AdS$
space. The metric of this spacetime is
\be 
ds^2/R^2_{ads} =  r^2 (d\tau^2+ dx^2_1 + dx^2_2 + dx^2_3) +
{ dr^2 \over r^2 } + d\Omega^2_5 \; , \nonumber
\ee
where the radius of the $AdS$ spacetime is given through $R^4_{AdS} =
g_s N \alpha'^2$ ($g_s$ is the string coupling and $l_s$ is the string
length, $l^2_s = \alpha'$). The Euclidean time is $\tau = i x_0$.  To
break conformal invariance, following ref. \cite{wittenT} , we place the
system at a nonzero temperature described by a periodic Euclidean time
$\tau = \tau + \beta$, $\beta=2 \pi R_0$. The metric correspondingly changes, 
for
small enough $R_0$, to the non-extremal black hole metric in $AdS$
space. For large black hole temperatures, the stable phase of the
metric corresponds to a black hole with radius large compared to the
$AdS$ curvature scale. To see the physics of discrete modes, we may
take the limit of going close to the horizon,
whereby the metric reduces to that of the black 3-brane. This metric
is, (where $
f(r)=r^2-{1\over r^2} \;
$, and we have scaled out all dimensionful quantities), 
\be
ds^2=f d\tau^2+f^{-1}dr^2+r^2(dx_1^1+dx_2^2+dx_3^2)+
d\Omega_5^2  \; ,
\label{eq:threep} 
\ee 

On the gauge theory side, we would have a ${\cal N}=4$ susy theory
corresponding to the $AdS$ spacetime, but with the $S^1$
compactification with antiperiodic boundary conditions for the
fermions, supersymmetry is broken and massless scalars are expected to
acquire quantum corrections. Consequently from the view point of a 3-d
theory, the compactification radius acts as an UV cut-off. Before the
compactification the 4-d theory was conformal, and was characterized
by a dimensionless effective coupling $({g_{YM}^{(4)}})^2N\sim g_s
N$. After the compactification the theory is not conformal, and the
radius of the compact circle provides a length scale. Let this radius
be $R_0$. Then a naive dimensional reduction from 4-d Yang-Mills to
3-d Yang-Mills, would give an effective coupling in the 3-d theory
equal to $(g_{YM}^{(3)})^2N=({g_{YM}^{(4)}})^2N/ (2 \pi R_0)$. The 3-d YM
coupling has the units of mass. If the dimensionless coupling
$({g_{YM}^{(4)}})^2N$ is much less than unity, then the length scale
associated with this mass is larger than the radius of compactification,
and we may expect the 3-d theory to be a dimensionally reduced version
of the 4-d theory. 

Unfortunately the dual supergravity description only applies at
$({g_{YM}^{(4)}})^2N>>1$, so that the higher Kaluza-Klein modes of the
$S^1$ compactification have lower energy than the mass scale set by
the 3-d coupling. Thus we do not really have a 3-d gauge theory with a
finite number of additional fields.  One may nevertheless expect that
some general properties of the dimensionally reduced theory might
survive the strong coupling limit. Moreover, we expect that the
pattern of spin splittings might be a good place to look for
similarities. In keeping with earlier work, we ignore the Kaluza-Klein
modes of the $S^1$ and restrict ourselves to modes that are singlets
of the $SO(6)$, since non-singlets under the $S^1$ and the $SO(6)$ can
have no counterparts in a dimensionally reduced $QCD_3$.

\section{Wave Equations}

We wish to consider fluctuations of the metric of the form, 
\be
g_{\mu\nu}=\bar g_{\mu\nu}+ h_{\mu\nu}(x)  \; ,
\ee
leading to the linear Einstein equation,
\be
h_{\mu\nu;\lambda}{}^{\lambda}
+h_\lambda^\lambda{}_{;\mu\nu}
-h_{\mu\lambda;\nu}{}^{\lambda}
-h_{\nu\lambda;\mu}{}^{\lambda} - 8  h_{\mu\nu}=0 \; . \nonumber
\label{eq:thir}
\ee
Our perturbations will have the form
\be
h_{\mu\nu}=\epsilon_{\mu\nu}(r)e^{-mx_3}
\ee
where we have chosen to use $x_3$ as a Euclidean time direction to
define  the glueball masses of the 3-d gauge theory. We fix  the 
gauge to $h_{3\mu}=0$.

>From the above ansatz and the metric, we see that we have an $SO(2)$
rotational symmetry in the $x_1-x_2$ space, and we can classify our
perturbations with respect spin.

{\bf Spin-2:} There are two linearly independent perturbations which 
form the spin-2 representation of $SO(2)$:
$h_{12}=h_{21}=q_T(r)e^{-mx_3}  ,  h_{11}=-h_{22}=q_T(r)e^{-mx_3} \;$
with {\rm all ~other~components~ zero}.  The Einstein equations give,
\be
(r^2-{1\over r^2})q_T'' + (r+{3\over r^3})q_T' +
({m^2\over r^2}-4 -{4\over r^4}) q_T = 0. \nonumber
\label{eq:fourt}
\ee
Defining $\phi_T(r)=q_T(r)/ r^2$,
this is the same equation as that satisfied  by the dilaton
(with constant value on the $S^5$).

{\bf  Spin-1: } The Einstein equation for the ansatz,
$h_{i\tau}=h_{\tau i}=q_V(r)e^{-mx_3}, {i=1 ,2}$
gives 
\be
(r^2-{1\over r^2})q_V'' +(r-{1\over r^3})q_V' +
({m^2\over r^2}-4 +{4\over r^4}) q_V =0 \; .
\ee

{\bf  Spin-0:} Based on the symmetries we choose an ansatz where the 
nonzero components of the perturbation are
\bea
h_{11} &=&h_{22}=q_1(r)e^{-mx_3}\nonumber\\
h_{\tau\tau}&=&-2q_1(r){f(r)\over
r^2}e^{-mx_3}+q_2(r)e^{-mx_3}\nonumber\\ h_{rr}&=&q_3(r)e^{-mx_3}
\nonumber
\eea
where $f(r)$ is defined above in the metric. The field equation for 
$q_3\equiv q_S(r)$, is
\be
p_2(r) q''_S(r) +   p_1(r) q'_S(r) +  p_0(r)  q_S(r) = 0,
\ee
where $  p_2(r) = r^2(r^4-1)^2 [ 3(r^4-1) + m^2 r^2 ]$, 
$p_1(r) = r(r^4-1)[3(r^4-1)(5r^4+3)+m^2r^2(7r^4+5)]$ and $p_0(r) =
9(r^4-1)^3+2m^2r^2(3+2r^4+3r^8)+m^4r^4(r^4-1)$.

\section{Numerical Solution}

To calculate the discrete spectrum for our three equation, one must 
apply the correct boundary conditions at $r = 1$ and $r = \infty$.
The result is a Sturm-Liouville problem for the propagation of 
gravitational fluctuations in a ``wave guide''.

{\bf Table 1. Glueball Excitation Spectrum}
\begin{table}[hbt]
\setlength{\tabcolsep}{1.0pc}
\newlength{\digitwidth} \settowidth{\digitwidth}{\rm 0}
\catcode`?=\active \def?{\kern\digitwidth}
\label{tab:levels}
\begin{tabular}{|l|r|r|r|} \hline
level & $ 0^{++}$ & $ 1^{-+}$  & $ 2^{++}$\\
\hline\hline
 n= 0   & 5.4573   & 18.676 & 11.588 \\ 
\hline
 n= 1   &  30.442  & 47.495 & 34.527 \\ 
\hline
 n= 2   & 65.123   & 87.722   & 68.975 \\ 
\hline
 n= 3   & 111.14     & 139.42  & 114.91 \\ 
\hline
 n= 4   & 168.60    & 203.99 & 172.33 \\ 
\hline
 n= 5   & 237.53   & 277.24 & 241.24 \\ 
\hline
 n= 6   & 317.93   & 363.38 & 321.63 \\ 
\hline
 n= 7   &  409.82  & 461.00  & 413.50 \\ 
\hline
 n= 8   & 513.18    &  570.11 & 516.86 \\ 
\hline
 n= 9   & 628.01   & 690.70 & 631.71 \\
\hline
\end{tabular}
\end{table}
\vskip -70mm
\hskip 75mm \psfig{figure=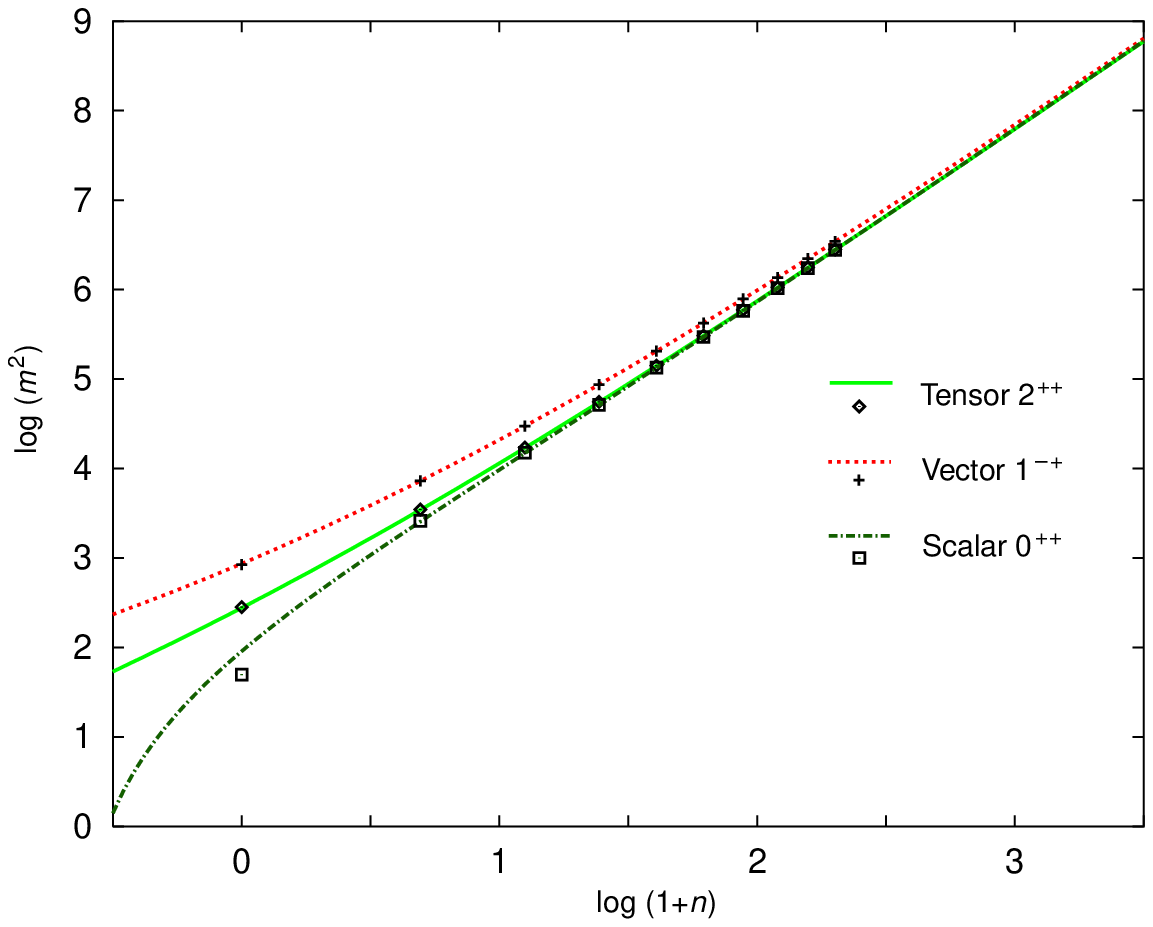,width=80mm,height=80mm,angle=0}

Using this shooting method we have computed the the first 10 states
given in Table \ref{tab:levels}.  The spin-2 equation is equivalent to
the dilaton equation \cite{csaki,jev}, so
the excellent agreement with earlier values validates our method.  
We used a standard Mathematica routine with
boundaries taken to be $x= r^2 -1 = \epsilon$ and $1/x = \epsilon$
reducing $\epsilon$ gradually to $\epsilon = 10^{-6}$.  Note that
since all our eigenfunctions must be even in $r$ with nodes spacing in
$x = r^2-1$ of $O(m^2)$, the variable $1/x$ is a natural way to
measure the distance to the boundary at infinity.  For both
boundaries, the values of $\epsilon$ was varied to demonstrate that
they were near enough to $r = 1, $ and $ \infty$ so as not to
substantially effect the answer.

As one sees in the accompanying figure, they match very accurately
with the leading order WKB approximation.
Simple variational forms also lead to very
accurate upper bounds for the ground state ($n = 0$) masses.

\section{Strong coupling Expansion for Pomeron Intercept}

Our current exercise has been extended to 4-d QCD using a scheme
involving the finite temperature version of $AdS^7 \times S^4$. As has
been suggested elsewhere, one goal is to find that background metric
that has the phenomenologically best strong coupling limit. This
should provide an optimal starting point for approaching the continuum
weak coupling regime.  Here, we shall report briefly the key 
constraint provided by the Pomeron intercept.

The Pomeron is the leading Regge trajectory passing through the
lightest glueball state with $J^{PC}=2^{++}$. In a linear
approximation, it can be parameterized by
\begin{equation}
\alpha_P(t) = 2 + {\alpha'_P} (t-m_T^2),
\end{equation}
where we can use the strong coupling estimate
for the lightest tensor mass\footnote{ We have adopted
the normalization in the $AdS$-black hole metric to simplify the
coefficients, e.g., for $AdS^7$, ${\bar g}_{\tau\tau}= r^2-
r^{-4}$. This corresponds to fixing the ``thermal-radius" $R_1=1/3$ so
that $\beta=2\pi R_1 =2\pi/3$.},
\begin{equation}
m_T \simeq [9.86 + 0( \frac{1}{g^2 N} )] \; \beta^{-1} \; .
\end{equation}
Moreover if we make the standard assumption that the closed string tension is
twice that between two static quark sources, we also have
a strong coupling expression for the  Pomeron slope,
\begin{equation}
\alpha_P'\simeq [ {27\over 32 \pi g^2 N}+ 0(\frac{1}{g^4 N^2})] \; { \beta^2}.
\end{equation}
Putting these together,
we obtain a strong coupling expansion for the Pomeron intercept,
\begin{equation}
\alpha_P(0) \simeq 2- 0.66 \; (\frac{4 \pi}{ g^2 N}) + 0(\frac{1}{g^4
N^2}) \; .
\end{equation}

Turning this argument around, we can estimate a crossover value between the
strong and weak coupling regimes by fixing $\alpha_P(0)
\simeq 1.2$ at its phenomenological value.  In fact this
yields for $QCD_4$ at $N = 3$ a reasonable value for $\alpha_{strong}
= g^2/4 \pi = 0.176$ for the  crossover. Much more experience with this new
approach to strong coupling must be gained before such numerology can
be taken seriously. However, similar crude argument have proven to be
a useful guide in the crossover regime of lattice QCD.  One might
even follow the general strategy used in the lattice cut-off
formulations. Postpone the difficult question of analytically solving
the QCD string to find the true UV fixed point.  Instead work at a
fixed but physically reasonable cut-off scale (or bare coupling) to
calculate the spectrum. If one is near enough to the fixed point, mass
ratios should be reliable. After all, the real benefit of a weak/strong
duality is to use each method in the domain where it provides the
natural language.  On the other hand, clearly from a fundamental point
of view, finding analytical tools to understand the renormalized
trajectory and prove asymptotic scaling within the context of the
gauge invariant QCD string would also be a major achievement --- an
achievement that presumably would include a proof of confinement
itself.

Results on these computations will be reported in a future
publication.\cite{bmt4}

{\noindent\bf Acknowledgments:} We would like to acknowledge useful
conversations with  R. Jaffe, A. Jevicki, D. Lowe, J. M. Maldacena, H.
Ooguri, and others.


\section*{References}
\begin{thebibliography}{99}
\bibitem{maldacena}J. Maldacena, Adv. Theor. Math. Phys. 2:231, 1998, 
hep-th/9711200.

\bibitem{wittenT}E. Witten, Adv. Theor. Math. Phys.2: 505, 1998, 
hep-th/9803131.

\bibitem{csaki} C. Cs\'aki, H. Ooguri, Y. Oz and J. Terning,
hep-th/9806021.

\bibitem{jev}R. De Mello Koch, A. Jevicki, M. Mihailescu and J. Nunes,
hep-th/9806125.

\bibitem{bmt}R. Brower, S. Mathur and C-I Tan, (to be published in 
Nucl. Physics),  hep-th/9908196.
\bibitem{cm}N. R. Constable and R.C. Meyers, hep-th/9908175.
\bibitem{bmt4}R. Brower, S. Mathur and C-I Tan, ``Glueball Spectrum for QCD from
AdS Supergravity Duality", hep-th/0003115.
\end{thebibliography}
\end{document}